\documentstyle[12pt]{article}

\begin{document}
\setcounter{page}{1} \pagestyle{plain} \vspace{1cm}
\begin{center}
\Large{\bf Failure of Standard Thermodynamics in Planck Scale Black Hole System }\\
\small
\vspace{1cm} {\bf Kourosh Nozari}\quad and \quad {\bf S. Hamid Mehdipour}\\
\vspace{0.5cm} {\it Department of Physics,
Faculty of Basic Sciences,\\
University of Mazandaran,\\
P. O. Box 47416-1467,
Babolsar, IRAN}\\
{\it e-mail: knozari@umz.ac.ir}

\end{center}
\vspace{1.5cm}

\begin{abstract}
The final stage of the black hole evaporation is a matter of debates
in the existing literature. In this paper, we consider this problem
within two alternative approaches: noncommutative geometry(NCG) and
the generalized uncertainty principle(GUP). We compare the results
of two scenarios to find a relation between parameters of these
approaches. Our results show some extraordinary thermodynamical
behavior for Planck size black hole evaporation. These extraordinary
behavior may reflect the need for a fractal nonextensive
thermodynamics for Planck size black hole evaporation process.
\end{abstract}
\newpage

\section{Introduction}
Usual uncertainty principle of Heisenberg should be modified to
incorporate quantum gravitational effects. Existence of a minimal
physical length ($\sim 10^{-33}cm$) which is a common feature of
alternative approaches to quantum gravity problem, restricts the
resolution of spacetime points[1,2]. This finite resolution of
spacetime can be addressed by the generalized uncertainty
principle(GUP). Consequences of GUP for various aspects of quantum
gravity problem have been studied extensively [3-7]. Among these
studies, black hole thermodynamics has found considerable
attentions[8-11]. Adler {\it et al} have argued that contrary to
standard viewpoint, GUP may prevent small black holes from total
evaporation in exactly the same manner that the uncertainty
principle prevents the Hydrogen atom from total collapse[12].
Embedding of black hole in a space-time with higher dimensions has
been studied in both compact and infinitely extended extra
dimensions[13]. The results of these
studies are summarized as follows\\
{\it Black hole evaporates by emission of Hawking radiation.  This
evaporation process terminates when black hole reaches a Planck size
remnant. This remnant has zero entropy, zero heat capacity and a finite
nonzero temperature.}\\
Of course, there are other alternative proposals to solve problem of
final stage of black hole evaporation such as approach based on
trace anomaly[14].\\
The idea of making spacetime noncommutative goes back to the early
days of quantum field theory, at least as early as 1947 [15]. The
idea was that considering a noncommutative structure of spacetime at
very small length scale, one could introduce an effective
ultraviolet cutoff. One of the main motivations is the hope that a
nontrivial structure of spacetime at small distances may led one to
quantum field theories with better ultraviolet behaviors. So, it
seems that both GUP and noncommutative geometry(NCG) provide
suitable framework for short distance behavior of physical systems.
Quantum correction of black hole thermodynamics within GUP and NCG
shows a similarity between the results of two approaches[16]. This
feature has its own importance since it can provide a better
understanding of the ultimate quantum gravity scenario. The purpose
of this paper is to consider the effect of space noncommutativity
and the generalized uncertainty principle on the short distance
thermodynamics of an evaporating Schwarzschild black hole. Our
analysis shows that extension of ordinary Boltzmann-Gibbs
thermodynamics to very short distance systems such as Planck size
black holes encounters severe difficulties. These difficulties may
reflect the need for a nonextensive thermodynamics such as Tsallis
thermodynamics[17]. As a possible connection between the results of
two scenarios, we compare our results obtained in NCG with the
results of GUP to find a relation between parameters of
corresponding theories.

\section{ Preliminaries}
The study of the structure of spacetime at Planck scale, where
quantum gravity effects are non-negligible, is one of the main open
challenges in fundamental physics. Since the dynamical variable in
Einstein general relativity is spacetime itself (with its metric
structure), and since in quantum mechanics and in quantum field
theory the classical dynamical variables should be noncommutative in
principle, one is strongly led to conclude that noncommutativity of
spacetime is a feature of Planck scale physics. This expectation is
further supported by Gedanken experiments that aim at probing
spacetime structure at very small distances. They show that due to
gravitational back reaction, one cannot test spacetime at Planck
scale. Its description as a (smooth) manifold becomes therefore a
mathematical assumption no more justified by physics. It is then
natural to relax this assumption and conceive a more general
noncommutative spacetime, where uncertainty relations and
discretization naturally arise. Noncommutativity is the central
mathematical concept expressing uncertainty in quantum mechanics,
where it applies to any pair of conjugate variables, such as
position and momentum. One can just as easily imagine that position
measurements might fail to commute and describe this using
noncommutativity of the coordinates.  The noncommutativity of
spacetime can be encoded in the commutator[18-20]
\begin{equation}
[\hat{x}^i,\hat{x}^j]=i\theta^{ij}
\end{equation}
where $\theta^{ij}$ is a real, antisymmetric and constant tensor,
which determines the fundamental cell discretization of spacetime
much in the same way as the Planck constant $\hbar$ discretizes the
phase space. In $d = 4$, by a choice of coordinates, this
noncommutativity can be brought to the form
\begin{displaymath}
\theta^{ij}= \left( \begin{array}{cccc}
0 & \theta & 0 & 0  \\
-\theta & 0 & \theta & 0 \\
0 & -\theta & 0 & \theta \\
0 & 0 & -\theta & 0
\end{array} \right)
\end{displaymath}
This was motivated by the need to control the divergences showing up
in theories such as quantum electrodynamics.  Although there has
been a long held belief that in theories of quantum gravity,
space-time must change its nature at distances comparable to the
Planck scale, but instead of trying to modify spacetime, focus was
directed to the fields defined on it. The outcome of these efforts
was what is known as string theory. The strings serve to smear out
the interaction in space-time and, in a sense, make the notion of a
point meaningless. There is a smallest distance that one can probe.
For this reason, in the context of string theories, this observable
distance is referred to generalized uncertainty principle- usual
uncertainty principle of quantum mechanics, the so-called Heisenberg
uncertainty principle, should be reformulated due to noncommutative
nature of spacetime at Planck scale. A GUP can be written as
follows[3]
\begin{equation}
\label{math:2.9}\Delta x\Delta p\geq\frac{1}{2}\Big(1+\alpha^2
l_p^2(\Delta p)^2\Big).
\end{equation}
Where $\alpha$ is a dimensionless and positive parameter of order
unity (for simplicity, we set $G=\hbar=c=1$). The main consequence
of  GUP is that measurement of the position is possible only up to
Planck length. So one can not setup a measurement to find more
accurate particle position than the Planck length, and this means
that the notion of locality breaks down. In other words, we cannot
look inside the region of minimal length. This minimal length
provides a natural cut off for underlying quantum field theory[21].
Based on this idea, it seems that the laws of physics should be
reformulated in very short distance systems. As we will show, the
need for a reformulation of the Planck size black hole
thermodynamics is inevitable.\\
After a brief overview of the conceptual preliminaries, we discuss
the issue of black hole thermodynamics in two alternative
approaches: GUP and NCG and finally we compare the results of these
two approaches. This comparison results an interesting relation
between parameters of these two scenario.

\section{Black Hole Thermodynamics with GUP}
In this section we derive black hole thermodynamics in GUP framework
based on our previous works[10,11,16]. Here we focus on mass
dependence of black hole thermodynamical properties. Using GUP as
our primary input, we obtain temperature, entropy and heat capacity
of a microscopically large Schwarzschild black hole. The results of
this calculations are interesting since they reflect some unusual
features of systems in very short distances.\\
In the current standard viewpoint, small black holes emit black body
radiation at the Hawking temperature. This temperature may be
obtained in a heuristic way with the use of the standard uncertainty
principle and general properties of black holes [22]. In this way,
we estimate the characteristic energy $E$ of the emitted photons
from the standard uncertainty principle. In the vicinity of the
black hole surface, there is an intrinsic uncertainty in the
position of any particle of about the Schwarzschild radius $r_s$,
due to the behavior of its field lines [23], as well as on
dimensional grounds. This leads to momentum uncertainty
\begin{equation}
\Delta p\approx\frac{1}{\Delta x}=\frac{1}{r_s}=\frac{1}{2M},\qquad
\Delta x\approx r_s=2M,
\end{equation}
and to an energy uncertainty of $\Delta E \approx \frac{1}{ 2M}$. We
identify this as the characteristic energy of the emitted photon,
and thus as a characteristic temperature; it agrees with the Hawking
temperature up to a factor of $4\pi $, which we will henceforth
include as a "calibration factor" and write, with $k_B = 1$,
\begin{equation}
T_{H}\approx \frac{1}{8\pi M},
\end{equation}
The related entropy is obtained by integration of\,\,
$dS=\frac{dM}{T}$\,\, which is the standard Bekenstein entropy,
\begin{equation}
S_{B}=4\pi M^2.
\end{equation}
However, if one consider the GUP as given by equation (2), the last
two equations become respectively,
\begin{equation}
T_{GUP}= \frac{M }{4\pi}\Bigg[1\mp \sqrt{1- \frac{1}{M^2}}\,\Bigg],
\end{equation}
and
\begin{equation}
S_{GUP}= 2\pi\Bigg[M^2+\sqrt{1- \frac{1}{M^2}}- \ln{\Big(M+\sqrt{M^2
- 1}\,\Big)}-1\Bigg].
\end{equation}
In equation (6), to recover the well-known results in the large mass
limit, one should consider the minus sign. As these equation show,
when the size of black hole approaches the Planck scale size, it
will cease radiation and its temperature reaches a maximum. This can
be seen from the behavior of the heat capacity. From equation (4),
we obtain standard heat capacity as follows
\begin{equation}
C_H=\frac{dM}{dT_H}=-8\pi M^2.
\end{equation}
If we consider  $T_{GUP}$ as given by equation (6), we obtain the
following generalized heat capacity
\begin{equation}
C_{GUP}=\frac{dM}{dT_{GUP}}=-\frac{4\pi\sqrt{1-\frac{1}{M^2}}}
{1-\sqrt{1-\frac{1}{M^2}}}.
\end{equation}
These equations strongly suggest the existence of black holes
remnants. As it is evident from figures 1, 2 and 3, in the framework
of GUP black hole can evaporate until when it reaches a remnant with
Planck mass. This remnant has zero entropy, zero heat capacity and a
non-zero maximal temperature. In the existing literatures there is
no obvious reason for this maximum temperature. As we will show,
within noncommutative geometry considerations, this maximum
temperature of remnant decreases and finally reaches to zero.
Vanishing of entropy for Planck size remnant seems to be strange and
needs more careful considerations.\\
One can show that within this view point black hole remnants are
stable[12].
\\
\\
\\
\\
\\

\begin{figure}[htp]
\begin{center}
\includegraphics{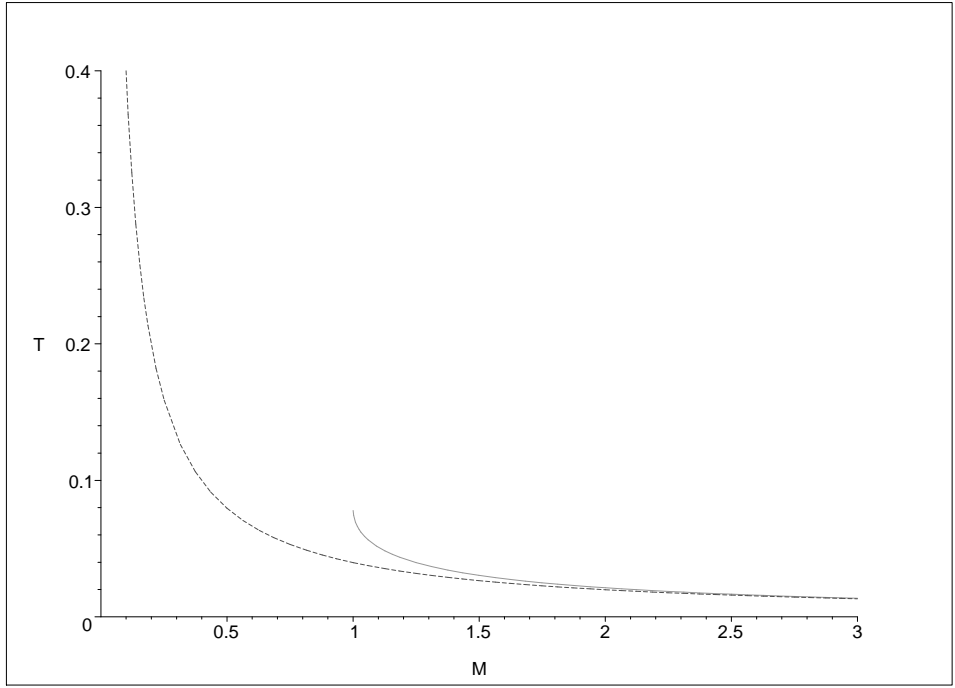}
\end{center}
\vspace{7.5 cm}
 \caption{\scriptsize {Temperature of a black hole versus its mass. Mass is in the
 units of Planck mass and temperature is in the units of the Planck energy.
 The lower curve (dashed line) is the well-known Hawking result,
 while the upper curve (line) is the result of GUP. }}
 \label{fig:1}
\end{figure}

\begin{figure}[htp]
\begin{center}
\includegraphics{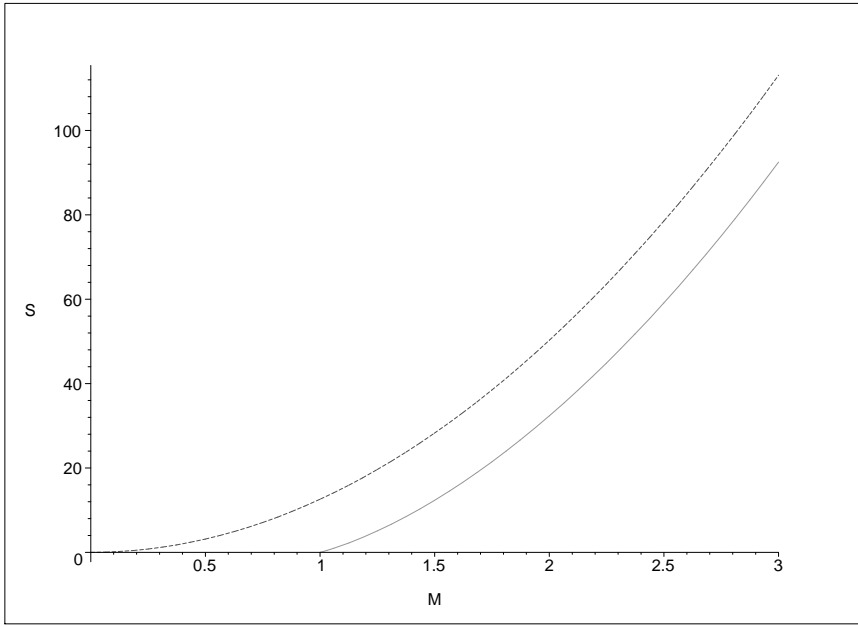}
\end{center}
\vspace{7.5 cm}
 \caption{\scriptsize { Entropy of a black hole versus its Mass. Entropy is dimensionless
 and mass is in the units of the Planck mass. The upper curve (dashed line) is
 the Hawking result, while the lower
 curve (line) is the result of GUP. }}
 \label{fig:1}
\end{figure}

\begin{figure}[htp]
\begin{center}
\includegraphics{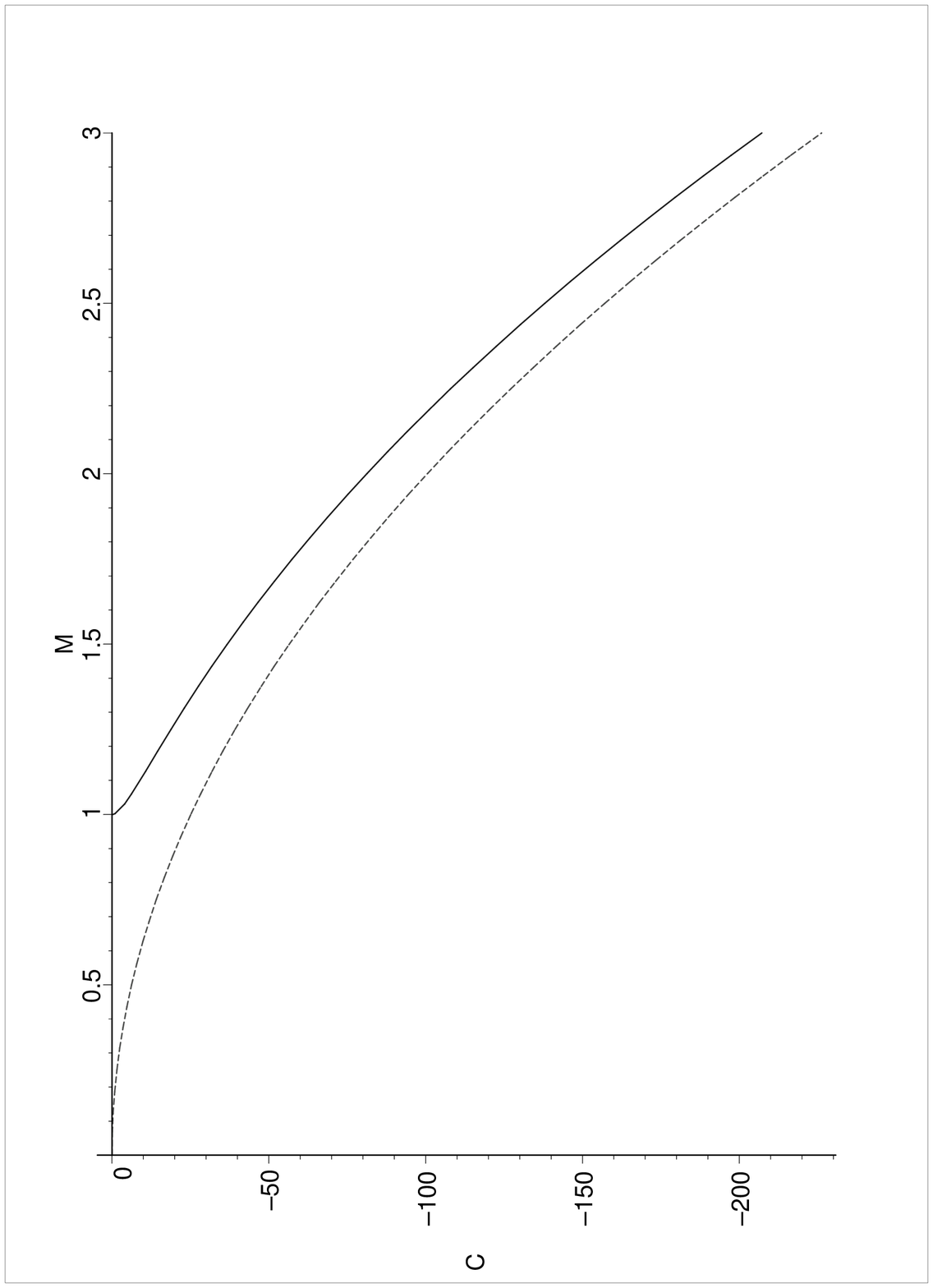}
\end{center}
\vspace{7.5 cm}
 \caption{\scriptsize{Heat capacity of a black hole versus the mass.
 The lower curve is the Hawking result (dashed line),
 while the upper curve (line) is the result of GUP.}}
 \label{fig:1}
\end{figure}

\section{Black Hole Thermodynamics in Noncommutative Spaces}
In this section we describe the effects of space noncommutativity on
the black hole thermodynamics.

There are two relatively different view point to incorporate space
noncommutativity in the issue of black hole thermodynamics. In one
of these view points, one considers the effect of space
noncommutativity on the radius of event horizon by a simple analysis
on coordinates noncommutativity[24,25]. In this framework one can
show that up to second order of noncommutativity parameter, the
noncommutative radius of event horizon has the following form[16]
\begin{equation}
{\hat r}_s =r_s -\frac{\zeta}{2 r_s}
+\frac{27}{8}\frac{\zeta^2}{r_s^{3}},\quad\quad\quad r_{s}=2M
\end{equation}
where
$$ \zeta = \frac{1}{16(1+\beta { p^{2}})^{2}} \left( p_x^2 + p_y^2
\right) \theta^2$$ and $p_{i}$ are components of black hole linear
momentum and $\beta$ is string theory parameter related to minimal
length[3]. Within this framework, perturbative thermodynamics of
noncommutative Schwarzschild black hole based on analysis presented
in [16] can be summarized as follows\\
Black hole temperature:
\begin{equation}
\hat T_{H}\approx \frac{1}{8\pi M}\Big[1+\frac{\zeta}{4 M^2}-\frac{3
\zeta^2}{8  M^4}\Big].
\end{equation}
Black hole entropy:
\begin{equation}
S\simeq \frac{A}{4}-\frac{\pi\zeta}{2} \ln{\frac{
A}{4}}+\sum_{n=1}^{\infty}c_{n}\Big(\frac{4}{A}\Big)^{n}+\cal{C}
\end{equation}
where \, $ \cal{C}$ is a constant of integration and $c_{n}$ are
constant. This framework considers the effect of noncommutativity on
geometric properties of black hole(such as black hole event horizon
radius) as starting point and one can obtain a modified
Schwarzschild line element using
modified event horizon radius.\\
There is another view point based on Nicolini {\it et al} works[26].
It has been shown [27] that noncommutativity eliminates point-like
structures in favor of smeared objects in flat spacetime. As
Nicolini {\it et al} have shown, the effect of smearing is
mathematically implemented as a substitution rule: position
Dirac-delta function is replaced everywhere with a Gaussian
distribution of minimal width $\sqrt{\theta}$. In this framework,
they have chosen the mass density of a static, spherically
symmetric, smeared, particle-like gravitational source as follows
\begin{equation}
\rho_\theta(r)=\frac{M}{(4\pi\theta)^{\frac{3}{2}}}\exp(-\frac{r^2}{4\theta})
\end{equation}
As they have indicated, the particle mass $M$, instead of being
perfectly localized at a point, is diffused throughout a region of
linear size $\sqrt{\theta}$. This is due to the intrinsic
uncertainty as has been shown in the coordinate commutator (1). This
matter source results the following static, spherically symmetric,
asymptotically Schwarzschild solution of the Einstein equations[26]
\begin{equation}
ds^2=\Bigg(1-\frac{4M}{r\sqrt{\pi}}\gamma\Big(\frac{3}{2},\frac{r^2}{4\theta}\Big)\Bigg)dt^2-
\Bigg(1-\frac{4M}{r\sqrt{\pi}}\gamma\Big(\frac{3}{2},\frac{r^2}{4\theta}\Big)\Bigg)^{-1}dr^2-r^2
(d\vartheta^2+sin^2\vartheta d\phi^2)
\end{equation}
where $\gamma\Big(\frac{3}{2},\frac{r^2}{4\theta}\Big)$ is the lower
incomplete Gamma function:
\begin{equation}
\gamma\Big(\frac{3}{2},\frac{r^2}{4\theta}\Big)\equiv\int_0^{\frac{r^2}{4\theta}}t^{\frac{1}{2}}e^{-t}dt
\end{equation}
The event horizon of this metric can be found where $g_{00} ( r_s )
= 0$,
\begin{equation}
r_s=\frac{4M}{\sqrt{\pi}}\gamma\Big(\frac{3}{2},\frac{r_s^2}{4\theta}\Big)
\end{equation}
As it is obvious from this equation, the effect of noncommutativity
in the large radius regime can be neglected, while at short distance
one expects significant changes due to the spacetime fuzziness. Now
the black hole temperature can be calculated as follows
\begin{equation}
T_{NCG}\equiv\Bigg(\frac{1}{4\pi\sqrt{-g_{00}g_{11}}}\frac{dg_{00}}{dr}\Bigg)_{r=r_s}=\frac{1}{4\pi
r_s}\Bigg[1-\frac{r_s^3}{4\theta^{\frac{3}{2}}}\frac{\exp\bigg(-\frac{r_s^2}{4\theta}\bigg)}
{\gamma\Big(\frac{3}{2},\frac{r_s^2}{4\theta}\Big)}\Bigg],
\end{equation}
where, $M$ has been expressed in terms of $r_s$ from the horizon
equation (16). For large black holes, where $\frac{r_s^2}{4\theta}
>> 1$, one recovers the standard result for the Hawking temperature
\begin{equation}
T_H=\frac{1}{4\pi r_s}
\end{equation}
At the initial stages of radiation, the black hole temperature
increases while the horizon radius is decreasing. It is important to
investigate what happens as $r_s \rightarrow\sqrt{\theta}$. In the
commutative case $T_H$, which is given by (18), diverges and this
puts limit on the validity of the conventional description of
Hawking radiation. Against this scenario, temperature (17) includes
noncommutative effects which are relevant at distances comparable to
$\sqrt{\theta}$[26].\\
Here, for numerical calculation purposes, for a moment we set
$\theta=1$, and rewrite equation (17) as follows

$$T_{NCG}=\frac{1}{8\pi
M\Big[1-\frac{2}{\sqrt{\pi}}\Gamma\Big(\frac{3}{2},M^2\Big)\Big]}\times$$
\begin{equation}
\Bigg[1-\frac{8M^3\Big[1-\frac{2}{\sqrt{\pi}}\Gamma\Big(\frac{3}{2},M^2\Big)\Big]^3
\exp\Big(-M^2\Big[1-\frac{2}{\sqrt{\pi}}\Gamma\Big(\frac{3}{2},M^2\Big)\Big]^2\Big)}
{4\gamma\Big(\frac{3}{2},M^2\Big[1-\frac{2}{\sqrt{\pi}}\Gamma\Big(\frac{3}{2},M^2\Big)\Big]^2
\Big)}\Bigg],
\end{equation}
where $\Gamma$ is upper Gamma function[26].  Behavior of the
noncommutative space temperature $T_{NCG}$ as a function of black
hole mass is plotted in figure $4$ and $5$ in two different limits.
\\
\\

\begin{figure}[htp]
\begin{center}
\includegraphics{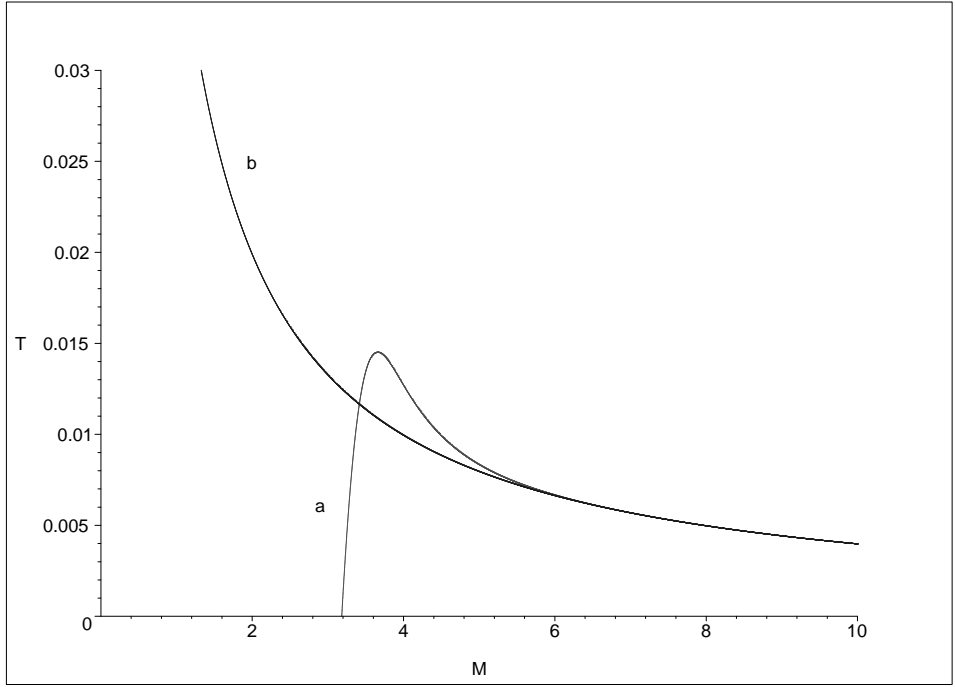}
\end{center}
\vspace{7.5 cm}
 \caption{\scriptsize  {Temperature of a black hole versus its mass. Mass is in the
 units of Planck mass and temperature is in the units of Planck energy.
 Curve(b) is the Hawking result,
 while curve (a) is the result of NCG.}}
 \label{fig:1}
\end{figure}

\begin{figure}[htp]
\begin{center}
\includegraphics{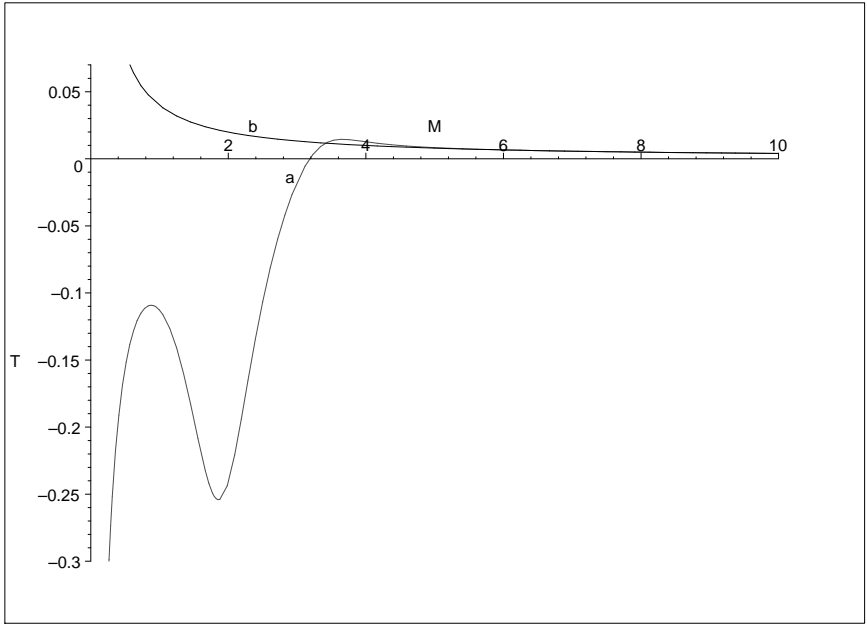}
\end{center}
\vspace{7.5 cm}
 \caption{\scriptsize  {Temperature of a black hole versus its mass. Mass is in the
 units of Planck mass and temperature is in the units of Planck energy. The lower
 curve(b) is the Hawking result,
 while the upper curve (a) is the result of NCG. In this figure we have considered
 the possibility of total evaporation.}}
 \label{fig:1}
\end{figure}

 As figure 4 shows, within noncommutative geometry,
temperature of black hole grows during its evaporation until it
reaches to a maximum extremal value and then falls down to zero. In
figure $5$ we have considered the possibility of complete
evaporation of black hole. If black hole evaporate completely,
noncommutative geometry consideration leads to a negative
temperature for black hole. We know from thermodynamics that
negative temperature can be reached by crossing high temperatures.
Therefore an extraordinary result is obtained for very short
distance system of Planck scale black hole. Note that Nicolini {\it
et al} have not considered the possibility of total evaporation.
They have plotted their figures for evaporation process which
continues only to a Planck size remnant[26]. They have not
considered the possibility of total evaporation. One point should be
stressed here: from GUP view point total evaporation is forbidden
while space noncommutativity consideration do not restrict
evaporation process to a Planck size remnant. In fact, equation of
black hole thermodynamics in noncommutative space allow the
possibility of total evaporation. For this reason we have considered
the possibility of total evaporation.\\
The entropy of the black hole can be obtained using the following
relation
\begin{equation}
S_{NCG}=\int_0^M\frac{dM}{T_{NCG}}.
\end{equation}
The numerical result of this integration is shown in figure $6$.
\\
\\

\begin{figure}[htp]
\begin{center}
\includegraphics{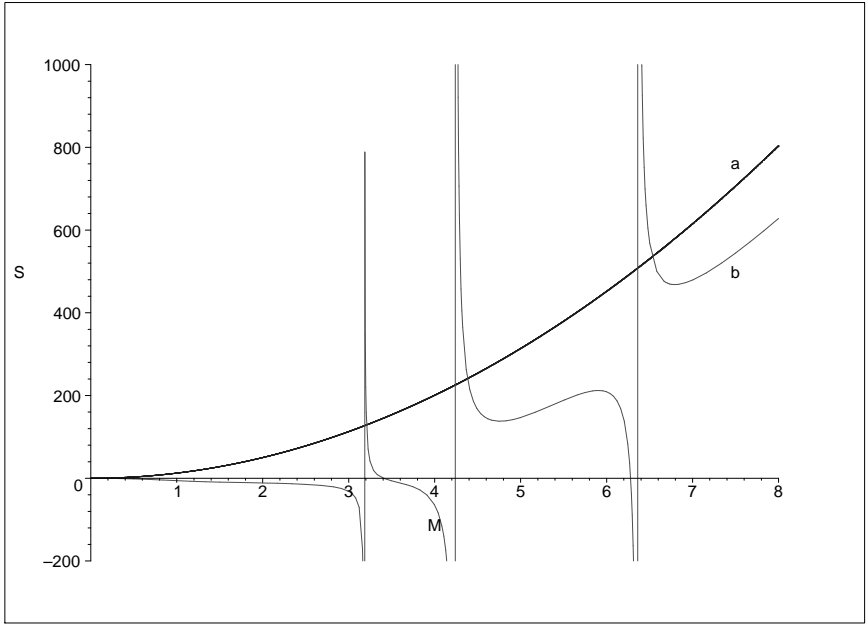}
\end{center}
\vspace{7.5 cm}
 \caption{\scriptsize  {Entropy of a black hole versus its mass. Entropy is dimensionless
 and mass is in the units of Planck mass. The upper curve (a) is the Hawking result, and the lower
 curve(b) is the result of NCG. This figure shows the failure of standard
 Boltzmann-Gibbs thermodynamics for Planck scale black holes.
 Negative entropy is a signature of this failure.}}
 \label{fig:1}
\end{figure}

As figure $6$ shows, entropy of black hole in noncommutative space
has some unusual behavior, specially it attains negative values for
some intervals of mass variation. This is physically meaningless.
This unusual thermodynamical behavior may be related to fractal
structure of spacetime in very short distances. In other words,
application of ordinary thermodynamics to situations such as Planck
scale black hole seems to be impossible. Due to non-extensive and
non-additive nature of such systems, one should apply non-extensive
formalism such as Tsallis thermodynamics[17]. Fractal nature of
spacetime at very short distances encourages the use of
non-extensive thermodynamics for Planck size black hole. The purpose
of this paper is to show the need for these non-extensive
thermodynamics for Planck size black holes. The formulation of such
a thermodynamics remains for future. We believe that theories such
as $E$-infinity[28] and scale relativity[29] which are based on
fractal structure of spacetime at very short distances, provide a
possible framework for thermodynamics of these short distance
systems. This
will be the subject of our forthcoming paper.\\
Heat capacity of black hole is given by
\begin{equation}
C_{NCG}=\bigg(\frac{dT_{NCG}}{dM}\bigg)^{-1},
\end{equation}
and its variation as a function of black hole mass is plotted in
figure $7$.
\\
\\

\begin{figure}[htp]
\begin{center}
\includegraphics{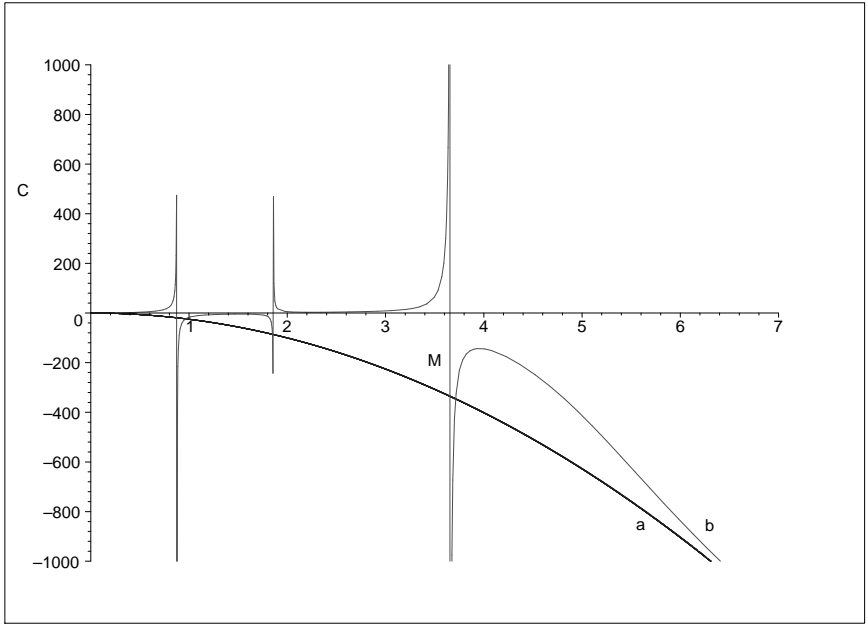}
\end{center}
\vspace{7.5 cm}
 \caption{\scriptsize  {Heat capacity of a black hole versus the mass.
 The lower curve (a), is the Hawking result(a) while the upper curve, (b)
 is the result of NCG.  Failure of standard thermodynamics is evident from this figure.}}
 \label{fig:1}
\end{figure}

\section{Extra Dimensional Considerations}
One surprising prediction of string theory is that several extra
dimension should be exist. This possible existence has opened up new
and exciting directions of research in quantum gravity. One of the
most significant sub-fields is the study of black hole production at
particle colliders, such as the Large Hadronic Collider (LHC)[30]
and the muon collider [31], as well as in ultrahigh energy cosmic
ray (UHECR) airshowers [32]. Furthermore, detection and vastly
production of such black holes at LHC could then be examined
experimentally in some details[33]. In such a scenario it would be
natural that GUP and NCG or both combination would now also be of
order $TeV$ allowing it to be accessible to colliders, because the
properties of such $TeV$-scale black hole may be influenced by NCG
and GUP or both combination effects, which originate at a similar
scale, and if these effects are large enough to be observable in
collider data output.\\
In a scenario with large extra dimensions(such as Arkani-Hamed,
Dimopoulos and Dvali(ADD) model[34]), GUP can be written as follows
\begin{equation}
\label{math:2.5} \Delta x_i\Delta p_i\geq
\frac{1}{2}\bigg(1+{\alpha}^2 L^2_{Pl}(\Delta
p_i)^2+\frac{{\beta}^2}{L^2_{Pl}}(\Delta x_i)^2+\gamma\bigg).
\end{equation}
Here $\alpha$, $\beta$ and $\gamma$ are dimensionless, positive and
independent of $\Delta x$ and $\Delta p$ but may in general depend
on the expectation values of $x$ and $p$. Planck length now is
defined as $ L_{Pl}= G_d^{\frac{1}{d-2}}$  where  $G_d$ is
gravitational constant in $d$-dimensional spacetime which in ADD
model is given by $G_{d} = G_{4}L^{d-4}$ ( $L$ is the extension of
the compactified dimensions). In which follows, we use this general
form of GUP as our primary input and construct a perturbational
calculations to find thermodynamical properties of black hole and
its quantum gravitational corrections. It should be stressed that
since GUP is a model independent concept, the results which we
obtain are consistent with any fundamental theory
of quantum gravity.\\
A $d$-dimensional spherically symmetric BH of mass $M$ (to which the
collider BHs will settle into before radiating) is described by the
metric[9],
\begin{equation}
\label{math:3.4}ds^2=-\bigg(1-\frac{16\pi
G_dM}{(d-2)\Omega_{d-2}r^{d-3}}\bigg)dt^2+\bigg(1-\frac{16\pi
G_dM}{(d-2)\Omega_{d-2}r^{d-3}}\bigg)^{-1}dr^2+r^2d\Omega^2_{d-2}
\end{equation}
where $\Omega_{d-2}$ is the metric of the unit $S^{d-2}$ as
$\Omega_{d-2}=\frac{2\pi^{\frac{d-1}{2}}}{\Gamma(\frac{d-1}{2})}$.
Since the Hawking radiation is a quantum process, the emitted quanta
should satisfy the generalized uncertainty principle(which has
quantum gravitational nature) in its general form. Therefore, we
consider equation (22), where $x_i$ and $p_i$ with $i = 1 . . . d -
1$, are the spatial coordinates and momenta respectively. By
modeling a BH as a $(d-1)$-dimensional cube of size equal to its
Schwarzschild radius $r_s$, the uncertainty in the position of a
Hawking particle at the emission is,
\begin{equation}
\label{math:3.4}\Delta x_i\approx
r_s=\omega_dL_{Pl}m^{\frac{1}{d-3}},
\end{equation}
where $$ \omega_d=\bigg(\frac{16\pi}{(d-2)\Omega_{d-2}}\bigg)
^{\frac{1}{d-3}},$$ $m=\frac{M}{M_{Pl}}$ and
$M_{Pl}=G_d^{-\frac{1}{d-2}}=L_{Pl}^{-1}$. Here $\omega_d$ is
dimensionless area factor. A simple calculation based on equation
(22) gives,
\begin{equation}
\label{math:3.1} \Delta x_i\simeq\frac{L_{Pl}^2 \Delta
p_i}{{\beta}^2 }\Bigg[1\pm\sqrt{1-{\beta}^2\bigg(
{\alpha}^2+\frac{(\gamma+1)}{L_{Pl}^2(\Delta
p_i)^2}\bigg)}\,\,\Bigg].
\end{equation}
where to achieve standard values (for example $\Delta x_i\Delta
p_i\geq 1$) in the limit of $\alpha=\beta=\gamma =0$, we should
consider the minus sign. One can minimize $\Delta x$ to find
\begin{equation}
\label{math:3.2}(\Delta x_i)_{min}\approx r_s(min)\simeq\pm \alpha
L_{Pl}\sqrt{\frac{1+\gamma}{1-\alpha^2\beta^2}}.
\end{equation}
This is the minimal observable length of the order of Planck length.
Here we should consider the plus sign whereas the negative sign has
no evident physical meaning. Equation (22) gives also
\begin{equation}
\label{math:3.3} \Delta p_i\simeq\frac{ \Delta
x_i}{{\alpha}^2L_{Pl}^2}\Bigg[1\pm\sqrt{1-{\alpha}^2\bigg({\beta}^2+\frac{
L_{Pl}^2(\gamma+1)}{(\Delta x_i)^2}\bigg)}\Bigg].
\end{equation}
To achieve correct limiting results we should consider the minus
sign in bracket. From a heuristic argument based on Heisenberg
uncertainty relation, one deduces the following equation for Hawking
temperature of black holes
\begin{equation}
\label{math:3.4}T_H\approx \frac{(d-3)\Delta p_i }{4\pi}
\end{equation}
where we have set the constant of proportionality equal to
$\frac{(d-3)}{4\pi}$ in extra dimensional scenario. Based on this
viewpoint, but now using generalized uncertainty principle in its
general form (22), modified black hole temperature in GUP is,
\begin{equation}
\label{math:3.5}T^{GUP}_{H}\approx \frac{(d-3) \Delta x_i}{{4\pi
\alpha}^2 L^2_{Pl}}\Bigg[1-\sqrt{1-{\alpha}^2\bigg(
{\beta}^2+\frac{L^2_{Pl}(\gamma+1)}{(\Delta
x_i)^2}\bigg)}\,\,\Bigg].
\end{equation}
Since $\Delta x_{i}$ is given by (24), this relation can be
expressed in terms of black hole mass in any stage of its
evaporation. Figure $8$ shows the relation between temperature and
mass of the black hole in different spacetime dimensions. Following
results can be obtained from this analysis : In scenarios with extra
dimensions, black hole has higher temperature. This feature leads to
faster decay and less classical behaviors for black holes. It is
evident that in extra dimensional scenarios final stage of
evaporation( black hole remnant) has mass more than its four
dimensional counterpart. Therefore, in the framework of GUP, it
seems that quantum black holes are hotter, shorter-lived and tend to
evaporate less than classical black holes. Note that these results
are applicable to both ADD and RS brane world scenarios.\\

\begin{figure}[htp]
\begin{center}
\includegraphics{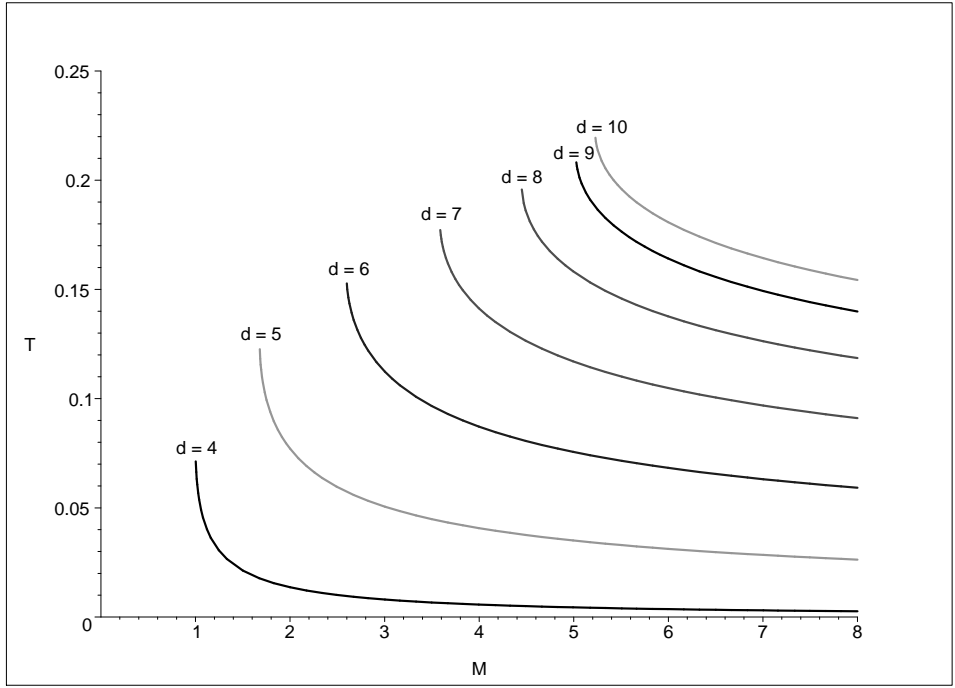}
\end{center}
\vspace{7.5 cm} \caption{\scriptsize {Temperature of black hole
Versus its mass in different spacetime dimensions. }} \label{fig:1}
\end{figure}

Now consider a quantum particle that starts out in the vicinity of
an event horizon and then ultimately absorbed by black hole. For a
black hole absorbing such a particle with energy $E$ and size $l$,
the minimal increase in the horizon area can be expressed as
\begin{equation}
\label{math:3.6}(\Delta \textsf{A})_{min}\geq\frac{ 8\pi
L_{Pl}^{d-2}E l}{(d-3)},
\end{equation}
then one can write
\begin{equation}
\label{math:3.7}(\Delta \textsf{A})_{min}\geq\frac{ 8\pi
L_{Pl}^{d-2}\Delta p_i l}{(d-3)},
\end{equation}
where $E\sim \Delta p_i $ and  $l\sim\Delta x_i$.
\begin{equation}
\label{math:3.8} (\Delta \textsf{A})_{min}\simeq\frac{8\pi
L_{Pl}^{d-4}(\Delta
x_i)^2}{(d-3){\alpha}^2}\Bigg[1-\sqrt{1-{\alpha}^2\bigg({\beta}^2+\frac{
L_{Pl}^2(\gamma+1)}{(\Delta x_i)^2}\bigg)}\Bigg],
\end{equation}
Now we should determine $\Delta x_i$. Since our goal is to compute
microcanonical entropy of a large black hole, near-horizon geometry
considerations suggests the use of inverse surface gravity or simply
the Schwarzschild radius for $\Delta x_i$. Therefore, $\Delta
x_i\approx r_s$ and defining $\Omega_{d-2} r_s^{d-2}=\textsf{A}$ or
$r_s^2=\Omega_{d-2}^{-\frac{2}{d-2}}\textsf{A}^{\frac{2}{d-2}}$ and
$(\Delta S)_{min}=b$, then it is easy to show that,
\begin{equation}
\label{math:3.8} (\Delta \textsf{A})_{min}\simeq\frac{8\pi
L_{Pl}^{d-4}\Omega_{d-2}^{-\frac{2}{d-2}}\textsf{A}^{\frac{2}{d-2}}}{(d-3){\alpha}^2}
\Bigg[1-\sqrt{1-{\alpha}^2\bigg({\beta}^2+\frac{
L_{Pl}^2(\gamma+1)}{\Omega_{d-2}^{-\frac{2}{d-2}}\textsf{A}^{\frac{2}{d-2}}}\bigg)}\Bigg],
\end{equation}
and,
\begin{equation}
\label{math:3.9}\frac{dS}{d\textsf{A}}\simeq\frac{(\Delta
S)_{min}}{(\Delta
\textsf{A})_{min}}\simeq\frac{\Omega_{d-2}^{\frac{2}{d-2}}b{\alpha}^2(d-3)}{8\pi
L_{Pl}^{d-4}\textsf{A}^{\frac{2}{d-2}}\Bigg[1-\sqrt{1-{\alpha}^2\bigg({\beta}^2+\frac{
\Omega_{d-2}^{\frac{2}{d-2}}L_{Pl}^2(\gamma+1)}{\textsf{A}^{\frac{2}{d-2}}}\bigg)}\Bigg]}.
\end{equation}
Note that $b$ can be considered as one bit of information since
within standard thermodynamics entropy is an extensive quantity.
Note also that in our approach we consider microcanonical ensemble
since we are dealing with Schwarzschild black hole of fixed mass.
Now we should perform integration. There are two possible choices
for lower limit of integration, $\textsf{A}=0$ and
$\textsf{A}=\textsf{A}_p$ . Existence of a minimal observable length
leads to existence of a minimum event horizon area, $\textsf{A}_p =
\Omega_{d-2} (\Delta x_i)_{min}^{d-2}$. So it is physically
reasonable to set $\textsf{A}_p$ as lower limit of integration. This
is in accordance with existing picture[12]. Based on these
arguments, we can write
\begin{equation}
\label{math:3.7}S\simeq\varepsilon\int_{\textsf{A}_{p}}^\textsf{A}\frac{\textsf{A}^{-\frac{2}{d-2}}}{1-\sqrt{\eta+\kappa
\textsf{A}^{-\frac{2}{d-2}}}}d\textsf{A}
\end{equation}
or\\
\begin{equation}
\label{math:3.7}S\simeq\varepsilon\int_{r_s(min)}^{r_s}\frac{(d-2)\Omega_{d-2}^{\frac{d-4}{d-2}}r_s^{d-5}
}{1-\sqrt{\eta+\kappa\Big(\Omega_{d-2}^{\frac{1}{d-2}}r_s\Big)^{-2}}}dr_s
\end{equation}

where, $$\varepsilon=
\frac{\Omega_{d-2}^{\frac{2}{d-2}}b\alpha^2(d-3)}{8\pi L_{Pl}^{d-4}
},\quad\quad \kappa=-{\Omega_{d-2}^{\frac{2}{d-2}}\alpha}^{2}
L_{Pl}^{2}(\gamma+1), \quad \quad \eta= 1-{\alpha}^{2}{\beta}^{2},$$
\begin{equation}
\label{math:3.7}\textsf{A}_{p}=\Omega_{d-2}\big(\alpha
L_{Pl}\big)^{d-2}\Bigg(\frac{1+\gamma}
{1-\alpha^2\beta^2}\Bigg)^{\frac{(d-2)}{2}}
\end{equation}
This integral can be solved numerically. The results are shown in
figure $9$. These figures show that: In scenarios with extra
dimensions, black hole entropy decreases. The classical picture
breaks down since the degrees of freedom of the black hole, i.e. its
entropy, is small. In this situation one can use the semiclassical
entropy to measure the validity of the semiclassical approximation.
It is evident that in extra dimensional scenarios final stage of
evaporation( black hole remnant) has event horizon area greater than
its four dimensional counterpart. Therefore, higher dimensional
black hole remnants have less classical features relative to their
four dimensional counterparts. In addition, as figure $9$ shows, for
large $d$ (for example $d\geq 8$), one find a linear area-entropy
relation but this linear entropy-area relation differs with standard
Bekenstein-Hawking result since it has greater slope. \\
\\
\\

\begin{figure}[htp]
\begin{center}
\includegraphics{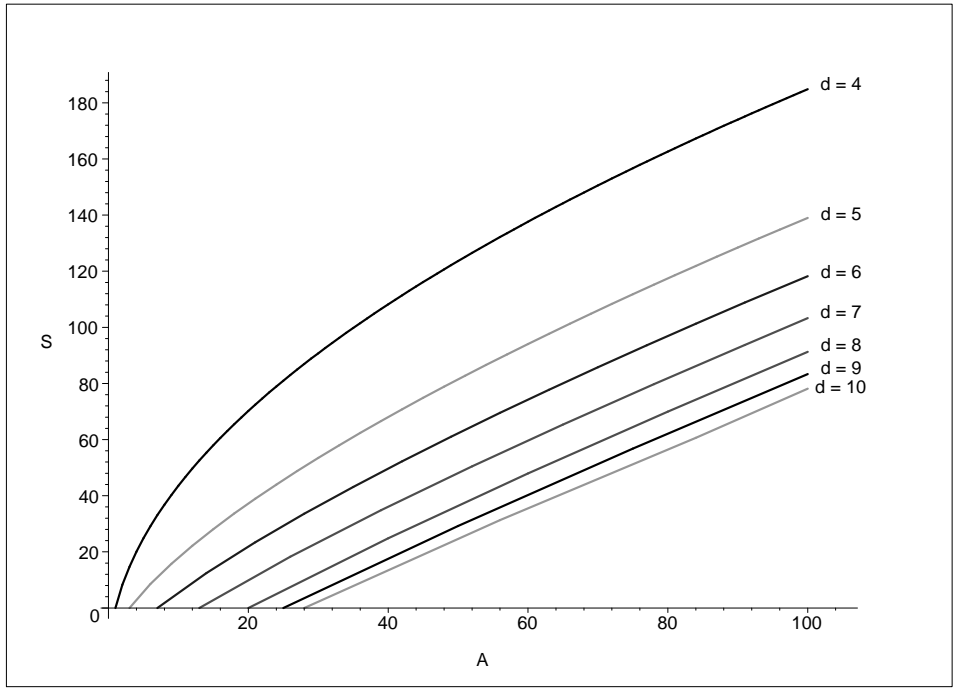}
\end{center}
\vspace{7.5 cm} \caption{\scriptsize {Entropy of black hole versus
the area of its event horizon in different spacetime dimensions. }}
\label{fig:1}
\end{figure}

Now, by considering noncommutative geometry in large extra
dimensional scenario, a static, spherically symmetric,
Gaussian-smeared matter source is given by[35]
\begin{equation}
\rho_{\theta,d}(r)={M\over {(4\pi \theta)^{\frac{(d-1)}{2}}}}
e^{-\frac{r^2}{4\theta}}.
\end{equation}
We assumed that the metric of our $d$-dimensional space is given by
\begin{equation}
ds^2=e^\nu dt^2-e^\mu dr^2-r^2 d\Omega_{d-2}^2.
\end{equation}
where
\begin{equation}
g_{00}=e^\nu=1-{1\over {M_{Pl}^{d-2}}} {M\over
{(d-2)\pi^{\frac{(d-1)}{2}}}} {1\over
{r^{d-3}}}\int_0^{\frac{r^2}{4\theta}} e^{-t}t^{\frac{(d-3)}{2}}dt.
\end{equation}
The horizon radius, $r_s$, occurs at values of $r$ where $g_{00}=0$.
Then, $r_s$ can be obtained by solving the equation
\begin{equation}
r_s=L_{Pl}\bigg[\frac{m}
{c_d}g_d\bigg(\frac{r_s}{\sqrt{\theta}}\bigg)\bigg]^{\frac{1}{d-3}},
\end{equation}
where $c_d={{(d-2) \pi^{\frac{(d-1)}{2}}}\over {\Gamma({{d-1}\over
{2}})}},$ $m=\frac{M}{M_{Pl}}$ and the functions
$g_d\big(\frac{r_s}{\sqrt{\theta}}\big)$, are given by the integrals
\begin{equation}
g_d\Big(\frac{r_s}{\sqrt{\theta}}\Big)={1\over \Gamma({{d-1}\over
{2}})}\int_0^{\frac{r_s^2}{4\theta}} e^{-t}t^{\frac{(d-3)}{2}}dt.
\end{equation}
Using arguments presented in preceding discussion about horizon area
and black hole entropy, relative integral can be written as
\begin{equation}
S\simeq\varepsilon\int_{r_s(min)}^{r_s}
\frac{(d-2)\Omega_{d-2}^{\frac{d-4}{d-2}}\varrho^{d-5}
g_d\big(\frac{r_s}{\sqrt{\theta}}\big)^{\frac{d-5}{d-3}}
}{1-\sqrt{\eta+\kappa\varrho^{-2}\Omega_{d-2}^{-\frac{2}{d-2}}
g_d\big(\frac{r_s}{\sqrt{\theta}}\big)^{-\frac{2}{d-3}}}}dr_s,
\end{equation}
where,
\begin{equation}
\varrho=\bigg(\frac{ML_{Pl}^{d-2}}{c_d}\bigg)^{\frac{1}{d-3}}
\end{equation}
Numerical calculation of this integral for different spacetime
dimensions are shown in figure $10$.
\\
\\

\begin{figure}[htp]
\begin{center}
\includegraphics{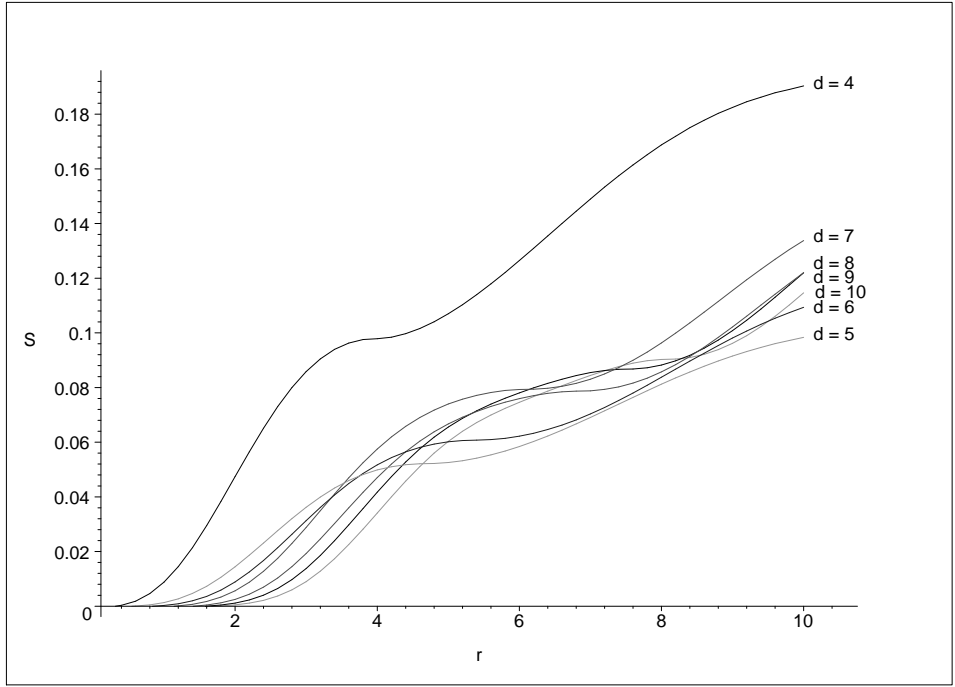}
\end{center}
\vspace{+8 cm} \caption{\scriptsize {Entropy of black hole versus
the radius of its event horizon in different spacetime dimensions
with considering both gravitational uncertainty and space's
noncommutativity. }} \label{fig:1}
\end{figure}

\section{The Relation Between Parameters of GUP and NCG }
Now we are going to compare the results of two approaches: GUP and
NCG. Using relation (2), we find
\begin{equation}
\Delta p\simeq\frac{\Delta
x}{\alpha^2l_p^2}\Bigg(1-\sqrt{1-\frac{\alpha^2l_p^2}{(\Delta
x)^2}}\Bigg).
\end{equation}
Within the original Bekenstein-Hawking framework and using the
previous results, one finds that there is a characteristic
temperature where agrees with the Hawking temperature up to a factor
of $4\pi$,
\begin{equation}
T_H\approx\frac{\Delta p}{4\pi},
\end{equation}
or
\begin{equation}
T_H^{GUP}\approx\frac{\Delta x}{4\pi
\alpha^2l_p^2}\Bigg(1-\sqrt{1-\frac{\alpha^2l_p^2}{(\Delta
x)^2}}\Bigg).
\end{equation}
for temperature of radiated photons in GUP framework. As relation
(2) shows, the position uncertainty has a minimum value of $(\Delta
x)_{min}=\alpha l_p$, so the string theory parameter of GUP times
the Planck distance play the role of a minimum or fundamental
distance which we take to be the Schwarzchild radius $r_s$. In the
region where $r_s=\alpha l_p$, $T_H^{GUP}$ deviates from the
standard hyperbola (18). Instead of exploding with shrinking $r_s$,
$T_H^{GUP}$ reaches a maximum,
\begin{equation}
T_H^{GUP}(max)=\frac{1}{4\pi \alpha l_p}.
\end{equation}
Comparing our two results (48)(with
$M|_{T_{(max)}}=3.63\sqrt{\theta}$) and (17), we find the following
relation between GUP and NCG parameters,
\begin{equation}
T_H^{GUP}(max)=T_H^{NCG}(max)\Longrightarrow\quad\frac{1}{4\pi
\alpha l_p}=\frac{0.0144}{\sqrt{\theta}}.
\end{equation}
So we find the following interesting relation between parameters of
GUP and NCG,
\begin{equation} \theta\simeq0.033\alpha^2l_p^2.
\end{equation}
Since $l_{p}\sim 10^{-33} cm$ and $\alpha\sim 1$, we find$$ \theta
\sim 10^{-68} cm^{2}.$$ In a model universe with spacelike extra
dimensions, we find the following generalization for $\Delta p_{i}$
\begin{equation}
\label{math:3.3} \Delta p_i\simeq\frac{ \Delta
x_i}{{\alpha}^2L_{Pl}^2}\Bigg[1-\sqrt{1-\frac{{\alpha}^2
L_{Pl}^2}{(\Delta x_i)^2}}\Bigg].
\end{equation}
Now GUP can be written as
\begin{equation}
\label{math:2.5} \Delta x_i\Delta p_i\geq
\frac{1}{2}\bigg(1+{\alpha}^2 L^2_{Pl}(\Delta p_i)^2\bigg),
\end{equation}
and we find the minimum of $\Delta x$ as follows
\begin{equation}
(\Delta x_i)_{min}\simeq\alpha L_{Pl}.
\end{equation}
Based on Heisenberg uncertainty relation, one deduces the following
equation for Hawking temperature of black holes in a universe model
with large extra dimensions
\begin{equation}
\label{math:3.4}T_H\approx \frac{(d-3)\Delta p_i }{4\pi}
\end{equation}
where we have set the constant of proportionality equal to
$\frac{(d-3)}{4\pi}$ in extra dimensional scenario. Therefore,
substitution of momentum uncertainty with the minimum value of
$\Delta x$, we obtain the maximum value for Hawking temperature,
\begin{equation}
T_H^{GUP}(max)\approx \frac{(d-3)}{4\pi \alpha L_{Pl}}
\end{equation}
In this extra dimensional scenario, the relation between parameters
of GUP and NCG is as follows
\begin{equation}
\theta\simeq\frac{0.033\alpha^2L_{Pl}^2}{(d-3)^2}.
\end{equation}

\section{Summary}
There are several alternative approaches for treating black hole
evaporation process. This process is a quantum gravitational effects
and its thorough understanding provides suitable framework toward a
complete formulation of quantum gravity proposal. In this paper,
based on our previous works and some recent literature, we have
discussed the final stage of black hole evaporation within two
alternative approaches: generalized uncertainty principle and the
space noncommutativity. Our calculations and the corresponding
enlightening figures show some unusual thermodynamical features when
the mass of the black hole becomes of the order of Planck mass or
less than it. Negative entropy, negative temperature, anomalous heat
capacity are some of these unusual features of this short distance
system. The origin of these unusual features may be on the failure
of standard thermodynamics at quantum gravity level. It seems that
standard formulation of thermodynamics breaks down at very short
distance systems. Due to fractal nature of spacetime at very short
distances, a new formulation of short distance thermodynamics is
inevitable. Theories such as $E$-infinity and scale relativity which
are based on fractal structure of spacetime at very short distances
may provide a suitable framework for formulation of this short
distance thermodynamics. The signature of this non-standard
thermodynamics has been seen in other problems such as very early
universe[36]. Currently there is no explicit formulation of such a
short distance formalism but in any case it should be based on
fractal nature of spacetime at quantum gravity level. As another
important result, we have compared temperature of black hole
calculated from two different viewpoint and we have found an
interesting relation between parameters of GUP and NCG. This
relation reveals the conceptual correspondence of GUP and NCG. We
are going to formulate fractal thermodynamics of Planck size black
hole in our forthcoming paper.


\begin{thebibliography}{10}
\bibitem{1}
- D. Amati, M. Ciafaloni and G. Veneziano, {\it Can spacetime be
probed below the string size?}, {\it Phys. Lett. B} {\bf 216} (1989)
41\\
-K. Konishi, G. Paffuti and P. Provero, {\it Minimum physical length
and the generalized uncertainty principle in string theory}, {\it
Phys. Lett. B} {\bf 234} (1990) 276
\bibitem{2}
L. J. Garay, {\it Quantum Gravity and Minimum Length},  {\it Int. J.
Mod. Phys. A} {\bf 10} (1995) 145
\bibitem{3}
A. Kempf {\it et al}, {\it Hilbert space representation of the
minimal length uncertainty relation}, {\it Phys. Rev. D} {\bf 52}
(1995) 1108-1118
\bibitem{4}
S. Hossenfelder {\it et al}, {\it Signature in the Planck Regime},
{\it Phys. Lett. B} {\bf 575} (2003) 85-99
\bibitem{5}
S. Hossenfelder, {\it Interpretation of Quantum Field Theories with
a Minimal Length Scale},  {\it Phys. Rev. D}  {\bf 73}  (2006)
105013
\bibitem{6}
K. Nozari and M. Karami, {\it Minimal length and the generalized
Dirac equation},  {\it Mod. Phys. Lett. A} {\bf 20} (2005) 3095-3104
\bibitem{7}
K. Nozari, {\it Some aspects of Planck scale quantum optics}, {\it
Phys. Lett. B} {\bf 629} (2005) 41-52
\bibitem{8}
- A. J. M. Medved and E. C. Vagenas, {\it When conceptual worlds
collide: The GUP and the BH entropy}, {\it Phys. Rev. D} {\bf 70}
(2004) 124021\\
- Y. S. Myung, Y. Kim and Y. Park, {\it Black hole thermodynamics
with generalized uncertainty principle}, [ arXiv:gr-qc/0609031]
\bibitem{9}
R. Casadio, B. Harms and Y. Leblanc, {\it Microfield Dynamics of
Black Holes}, {\it Phys.Rev. D} {\bf 58} (1998) 044014
\bibitem{10}
K. Nozari and S. Hamid Mehdipour, [ arXiv:gr-qc/0511110], {\it Int.
J. Mod. Phys. A}  (In Press)
\bibitem{11}
K. Nozari and S. H. Mehdipour, {\it Mod. Phys. Lett. A} {\bf 20}
(2005) 2937-2948
\bibitem{12}
R. J. Adler, P. Chen, D. I. Santiago, {\it The Generalized
Uncertainty Principle and Black Hole Remnants},  {\it Gen. Rel.
Grav.} {\bf 33} (2001) 2101.
\bibitem{13}
- P. C. Argyres, S. Dimopoulos and J. March-Russell, {\it Black
Holes and Sub-millimeter Dimensions }, {\it Phys. Lett. B} {\bf 441}
(1998) 96-104.\\
- A. Chamblin, S. Hawking and H.S. Reall, {\it Brane-World Black
Holes}, {\it Phys. Rev. D} {\bf 61} (2000) 065007.
\bibitem{14}
- R. Casadio, {\it Holography and trace anomaly: what is the fate of
(brane-world) black holes?}, {\it Phys.Rev. D} {\bf 69} (2004)
084025\\
- R. Casadio and C. Germani, {\it Gravitational collapse and black
hole evolution: do holographic black holes eventually
"anti-evaporate"?} {\it Prog.Theor.Phys.} {\bf 114} (2005) 23-56
\bibitem{15}
H. S. Snyder, {\it Quantized Spacetime}, {\it Phys. Rev.}  {\bf 71}
(1947) 38.
\bibitem{16}
K. Nozari and B. Fazlpour, [arXiv:hep-th/0605109]  and
[arXiv:gr-qc/0608077]
\bibitem{17}
- C. Tsallis, {\it Nonextensive statistics: Theoretical,
experimental and computational evidences and connections},  {\it
Braz. J. Phys.} {\bf 29} (1999)1-35\\
- C. Tsallis {\it et al},  {\it  Introduction to Nonextensive
Statistical Mechanics and Thermodynamics}, [arXiv:cond-mat/0309093]\\
- C. Tsallis and E. Brigatti, {\it Nonextensive statistical
mechanics: A brief introduction}, {\it Continuum  Mech. Thermodyn.}
{\bf 16} (2004) 223-235
\bibitem{18}
J. Madore, {\it An Introduction to Noncommutative Differential
Geometry and its Applications}, Cambridge University Press,
Cambridge, 1995.
\bibitem{19}
N. Seiberg and E. Witten,  {\it JHEP} {\bf 9909} (1999) 032
\bibitem{20}
M. R. Douglas and N. A. Nekrasov, {\it Noncommutative Field Theory},
{\it Rev. Mod. Phys.} {\bf 73} (2001) 977-1029
\bibitem{21}
S. Hossenfelder, {\it Running coupling with minimal length}, {\it
Phys. Rev. D} {\bf 70} (2004) 105003
\bibitem{22}
H. Ohanian and R. Ruffini, {\it Gravitation and Spacetime}, Second
Edition, W. W. Norton (1994),  p. 481.
\bibitem{23}
- R. J. Adler and T. K. Das, {\it Phys. Rev. D} {\bf 14} (1976) 2472\\
- R. S. Hanni and R. Ruffini, {\it Lines of Force of a Point Charge
Near a Schwarzschild Black Hole}, in {\it Black Holes}, Editted by
C. DeWitt and B. S. Dewitt (Gordon Breach, 1973).
\bibitem{24}
F. Nasseri, {\it Event Horizon of Schwarzschild Black Hole in
Noncommutative Spaces}, {\it Int. J. Mod. Phys. D} {\bf15} (2006)
1113-1118
\bibitem{25}
M. R. Setare, {\it Space Noncommutativity Corrections to the
Cardy-Verlinde Formula}, {\it Int. J. Mod. Phys. A}  {\bf21} (2006)
3007-3014
\bibitem{26}
- P. Nicolini {\it et al.}, {\it Noncommutative geometry inspired
Schwarzschild black hole}, {\it  Phys. Lett. B} {\bf 632} (2006)
547\\
- P. Nicolini, {\it A model of radiating black hole in
noncommutative geometry}, {\it J. Phys. A} {\bf 38} (2005) L631-L638
\bibitem{27}
- A. Smailagic and E. Spallucci, {\it Feynman Path Integral on the
Noncommutative Plane },  {\it J. Phys. A} {\bf 36} (2003) L467\\
- A. Smailagic and  E. Spallucci, {\it UV divergence-free QFT on
noncommutative plane }, {\it  J. Phys. A} {\bf 36} (2003) L517
\bibitem{28}
- M. S. El Naschie, The concepts of E-infinity: An elementary
introduction to the Cantorian-fractal theory of quantum physics,
{\it Chaos, Solitons and Fractals} {\bf 22} (2004) 495-511\\
- M. S. El Naschie,  Elementary prerequisites for
E-infinity(Recommended background readings in nonlinear dynamics,
geometry and topology), {\it Chaos, Solitons and Fractals}, {\bf 30}
(2006) 579-605\\
- M. S. El Naschie, Intermediate prerequisites for
E-infinity(Further recommended reading in nonlinear dynamics and
mathematical physics), {\it Chaos, Solitons and Fractals}, {\bf 30}
(2006) 622-628\\
- M. S. El Naschie, Advanced prerequisites for E-infinity theory,
{\it Chaos, Solitons and Fractals}, {\bf 30} (2006) 636-641
\bibitem{29}
L. Nottale, {\it Fractal space-time and microphysics: towards a
theory of scale relativity}, World Scientific, Singapore, 1993
\bibitem{30}
- http://lhc-new-homepage.web.cern.ch/lhc-new-homepage/\\
-  B. Koch, M. Bleicher and S. Hossenfelder, {\it Black Hole
Remnants at the LHC}, {\it JHEP 0510 (2005) 053}\\
-  S. Hossenfelder, B. Koch and M. Bleicher, {\it Trapping Black
Hole
Remnants}, [arXiv:hep-ph/0507140]\\
- T. J. Humanic, B. Koch and H. Stoecker, {\it Signatures for Black
Hole production from hadronic observables at the Large Hadron
Collider}, [ arXiv:hep-ph/0607097], {\it Int. J. Mod. Phys. E}  (In
Press)
\bibitem{31}
- http://www.fnal.gov/projects/muon collider/\\
- R. Godang, S. Bracker, M. Cavaglia, L. Cremaldi, D. Summers and D.
Cline, {\it Resolution of Nearly Mass Degenerate Higgs Bosons and
Production of Black Hole Systems of Known Mass at a Muon Collider },
{\it Int.J.Mod.Phys. A}, {\bf 20} (2005) 3409-3412
\bibitem{32}
- J. L. Feng and A. D. Shapere, {\it Black Hole Production by Cosmic
Rays }, {\it Phys. Rev. Lett.} {\bf 88} (2002) 021303\\
- A. Ringwald and H. Tu, {\it Phys. Lett. B} {\bf 525} (2002) 135
[arXiv:hep-ph/0111042]; {\it Phys. Lett. B} {\bf 529} (2002) 1
[arXiv:hep-ph/0201139]\\
- M. Ave, E. J. Ahn, M. Cavagli\`a and A. V. Olinto, {\it Phys. Rev.
D} {\bf 68} (2003)  043004  [arXiv:astro-ph/0306344]
\bibitem{33}
- M. Cavagli\`a  and S. Das, {\it How classical are TeV-scale black
holes?},  {\it Class. Quant. Grav.} {\bf 21} (2004) 4511-4522\\
- M. Cavagli\`a, {\it Int. J. Mod. Phys. A} {\bf18}, (2003) 1843
[arXiv:hep-ph/0210296]\\
- M. Cavagli\`a, Saurya Das and  R. Maartens, {\it Class. Quant.
Grav.} {\bf 20} (2003) L205-L212 [arXiv:hep-ph/0305223]\\
- G. Landsberg, {\it Eur. Phys. J. C} {\bf 33} (2004) S927-S931,
[arXiv:hep-ex/0310034]; {\it J. Phys. G} {\bf 32} (2006) R337-R365,
[arXiv:hep-ph/0607297]\\
- P. Kanti, {\it Int. J. Mod. Phys. A} {\bf 19} (2004)
4899-4951 [arXiv:hep-ph/0402168]\\
- M. Cavagli\`a, R. Godang, L. Cremaldi, D. Summers,
[arXiv:hep-ph/0609001]
\bibitem{34}
N. Arkani-Hamed, S. Dimopoulos and G. Dvali, {\it Phys. Lett. B}
{\bf 429} (1998)  263-272 [arXiv:hep-ph/9803315]; I. Antoniadis, N.
Arkani-Hamed, S. Dimopoulos and G. Dvali, {\it Phys. Lett. B} {\bf
436} (1998)  257-263 [arXiv:hep-ph/ 9804398]; N. Arkani-Hamed, S.
Dimopoulos and G. Dvali, {\it Phys. Rev. D} {\bf 59} (1999) 086004
[arXiv:hep-ph/9807344]
\bibitem{35}
T. G. Rizzo, {\it Noncommutative Inspired Black Holes in Extra
Dimensions }, [arXiv:hep-ph/0606051]
\bibitem{36}
K. Nozari and B. Fazlpour, {\it Generalized Uncertainty Principle,
Modified Dispersion Relations and Early Universe Thermodynamics}
[arXiv:gr-qc/0601092] , to appear in General Relativity and
Gravitation.


\end{thebibliography}
\end{document}